\documentclass[aps,prl,twocolumn,superscriptaddress]{revtex4}
\usepackage[usenames]{color}
\newcommand{\comment}[1]{}
\usepackage{graphicx}
\usepackage{amsmath}
\usepackage{epsfig}

\def\be{\begin{eqnarray}}
\def\ee{\end{eqnarray}}

\def\l{\left}
\def\r{\right}
\def\yd{^\dagger}
\def\cl{{\cal L}}

\def\bk{{\bf k}}
\def\bq{{\bf q}}

\begin{document}

\bibliographystyle{plain}
\title{Mott transition between a spin-liquid insulator and a metal
in three dimensions}

\date{\today}

\author{Daniel Podolsky}
\affiliation{Department of Physics, University of Toronto, Toronto,
Ontario M5S 1A7 Canada}
\author{Arun Paramekanti}
\affiliation{Department of Physics, University of Toronto, Toronto,
Ontario M5S 1A7 Canada}
\author{Yong Baek Kim}
\affiliation{Department of Physics, University of Toronto, Toronto,
Ontario M5S 1A7 Canada}
\author{T. Senthil}
\affiliation{Department of Physics, Massachusetts Institute of Technology,
Cambridge, Massachusetts 02139, USA}

\begin{abstract}
We study a bandwidth controlled Mott metal-insulator transition (MIT)
between a Fermi liquid metal and a quantum spin-liquid insulator at
half-filling in three dimensions (3D). Using a slave rotor approach,
and incorporating gauge field fluctuations, we find a continuous MIT and
discuss the finite temperature crossovers around this critical point.
We show that the specific heat $C \sim T \ln \ln (1/T)$
at the MIT and argue that the electrical transport on the
metallic side near the transition should exhibit a `conductivity minimum'
as a function of temperature.
A possible candidate to test these predictions is the 3D spin liquid
insulator Na$_4$Ir$_3$O$_8$ which exhibits a pressure-tuned transition
into a metallic phase. We also present the electron spectral function
of Na$_4$Ir$_3$O$_8$ at the transition.
\end{abstract}

\maketitle

{\it Introduction.---} 
Recent experiments suggest that Na$_4$Ir$_3$O$_8$ at
ambient pressure may be the first example of a three-dimensional quantum 
spin-liquid Mott-insulator\cite{okamoto}. This material has Ir local
moments residing on the hyper-kagome lattice which is a three-dimensional
(3D) network formed by corner sharing triangles.
Of the various theoretical proposals for the spin liquid phase observed
in this material
\cite{hopkinson07,lawler08a,lawler08b,palee08,chen08}, the data may be
most consistent with a particular gapless
spin liquid whose low energy excitations are charge-neutral 
spin-1/2 fermions, called spinons, which 
live on `spinon Fermi surfaces' in momentum space \cite{lawler08b,palee08}.
Remarkably, very recent experiments show that this material undergoes a  
transition to a metallic state under pressure \cite{takagi08}.
This indicates that Na$_4$Ir$_3$O$_8$ may be a 3D
analog of the well-studied layered triangular lattice organic material 
$\kappa-$(ET)$_2$Cu$_2$(CN)$_3$ \cite{kanoda03} which is insulating at
ambient pressure but undergoes a Mott transition to 
a metal under moderate pressure. 
Theoretically, the proximity to the Mott transition 
has been argued to lead to a 2D
spin liquid with a `spinon Fermi surface'
in $\kappa-$(ET)$_2$Cu$_2$(CN)$_3$ \cite{motrunich05,leelee05}.

\begin{figure}
\includegraphics[width=2.8in]{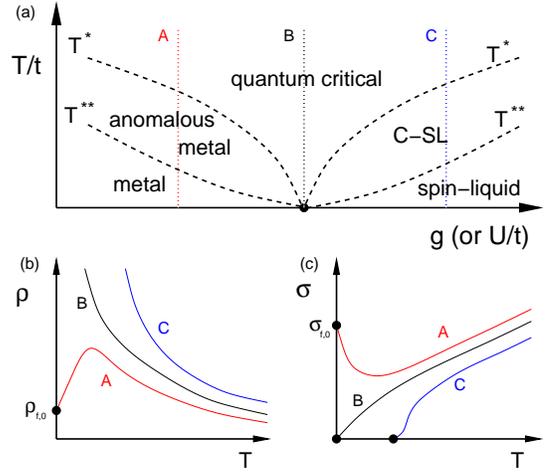}
\caption{{\bf (a)} Schematic phase diagram of the 3D metal insulator transition.
Finite temperature crossovers at the temperatures $T^*$
and $T^{**},$ shown by dashed lines, separate
regions distinguished by the specific heat and
conductivity behaviors, as discussed in the text
(``C-SL'' labels the Charge fluctuation
renormalized Spin Liquid).
{\bf (b)} Resistivity and {\bf (c)} conductivity along the lines -- A, B, and C -- 
in panel {\bf (a)}.
\label{fig:mf} } \vskip-0.2in
\end{figure}

In this Letter, motivated by the recent experiments on 
Na$_4$Ir$_3$O$_8$, we present a theory of a continuous Mott transition between 
a spin-liquid with a spinon Fermi surface and metal in 3D.
We use a similar theoretical framework as developed earlier by one of us\cite{senthil2d} for
a 2D Mott transition between a Fermi liquid and a spin liquid 
Mott insulator
in relation to $\kappa-$(ET)$_2$Cu$_2$(CN)$_3$. In 2D,
several interesting results such as a universal jump in the residual resistivity and the emergence of marginal Fermi liquids in an intermediate crossover scale were found.
In this work, we show that a continuous Mott transition is indeed possible in 
3D
and moreover it is characterized by weaker singularities in various physical 
quantities when 
compared to the 2D case. In contrast to 2D, right at the Mott critical point
in 3D, the system is insulating. Indeed we argue that the electrical conductivity $\sigma (T)$ in the vicinity of the Mott quantum critical point (QCP)
should exhibit the following
behavior: (i) In the `quantum critical' (QC) regime, $T > T^{*}$, we expect $\sigma (T)
\sim T$ (with log corrections). Here the crossover scale $T^* \sim |g - g_c|^{\frac{1}{2}}$ where $g$ is the parameter used to tune the transition (for instance the pressure, or the strength of the Coulomb repulsion $U/t$). $g_c$ is the value of the parameter at the zero temperature Mott QCP.
(ii) On the insulating side, for $T < T^{*}$,
the conductivity is thermally activated.  (iii) Finally, on the
metallic side, for $T \sim T^{*}$, we argue that $\sigma(T)$
will have a `conductivity minimum' before it grows and
saturates at low temperature. Equivalently there will be a resistivity peak at a non-zero temperature. This is sketched in Fig.~\ref{fig:mf}(b) and (c).  This resistivity peak is the most striking result of this paper.

As in the 2D case the finite temperature crossovers near the Mott transition are characterized by multiple energy scales $T^*$ and $T^{**}$  as depicted schematically in Fig.~\ref{fig:mf}(a). As noted above the scale $T^*$ determines the crossover in transport properties associated with the Mott transition.
However there are only weak changes in physical properties across the scale $T^{**} \sim |g-g_c|^{\frac{3}{2}}$.
For instance for $T \gg T^{**},$ in the regimes labelled quantum critical
(QC), `anomalous metal' and `charge fluctuation renormalized spin liquid'
(C-SL), the specific heat behaves as $C \sim T \ln\ln (1/T)$. For $T \ll T^{**}$ in the Mott insulator
$C \sim T \ln(1/T)$ while in the same regime on the metallic side we
recover the Fermi liquid result
$C \sim T$.
With an eye toward application to Na$_4$Ir$_3$O$_8$, we also present
predictions for the $T=0$
electron spectral function and tunneling density of states
at the Mott transition for a tight binding model appropriate to this 
material.

{\it Microscopic Model and Field Theory.---} To understand the universal features of this kind of 3D Mott transition we start with a single band Hubbard 
model at half-filling on a non-bipartite lattice,
$H=-t\sum_{\langle ij\rangle}c_{i\sigma}\yd c_{j\sigma}+\frac{U}{2}
\sum_i n_i^2-\mu\sum_i n_i,$
where $c_{i\sigma}\yd$ is the creation operator for the electron at site $i$
with
spin $\sigma$, and $n_i = c^{\dagger}_{i\sigma} c_{i \sigma}$ is the electron
number operator (repeated spin indices are summed over).
This model Hamiltonian is expected to have a metallic
Fermi-liquid ground state for small on-site repulsion
$U/t \ll 1$, and to have a Mott insulating ground state
for $U/t \gg 1$. (Henceforth, we set $t\!=\!1$.) The arguments of Ref. \onlinecite{motrunich05}
suggest that a spin liquid state with a spinon Fermi surface is a likely ground state just on the
insulating side of Mott transition.

The Mott transition between a Fermi-liquid metal and such a spin-liquid
insulator is most conveniently
accessed in a slave rotor formalism \cite{florens04},
in which the electron operator
is written as a product, $c_{i \sigma}\yd={\rm e}^{i\theta_i}f_{i\sigma}\yd$,
of a charged spinless `rotor'
field $\phi_i = {\rm e}^{i\theta_i}$ and a spin-1/2 charge-neutral
`spinon' $f_{i\sigma}$. In order to eliminate unphysical states from
the enlarged spinon-rotor Hilbert space, we must impose the constraint
$n^{\theta}_i + n^f_i = 1$ at each
site. Upon coarse-graining, this theory can be recast in the form
of a `gauge theory' which consists of a dynamical U(1) gauge field
$a_\mu = (a_0, {\textbf {\textit a}})$ coupled to, both, the spinon
and rotor fields \cite{leelee05}.
We begin by
writing down a continuum theory to describe this physics in
terms of a complex bosonic
field $\phi \sim e^{i \theta}$, a fermionic field $\psi_\sigma$, and a gauge
field $a_\mu$, with action $S=\int_0^\beta d\tau\int d^3{\bf r} \,{\cl}$, where
\begin{eqnarray}
\cl&=&\cl_b+\cl_f+\cl_g+\cl_{bf} \cr
\cl_b&=&\left|(\partial_\mu+ia_\mu)\phi\right|^2+m^2|\phi|^2+ u |\phi|^4\cr
\cl_f&=&\psi_\sigma\yd (\partial_\tau-ia_0-\mu_f)\psi_\sigma+\frac{1}{2m_f}\left|(\nabla-i {\textbf {\textit a}})\psi_\sigma\right|^2\cr
\cl_g&=&\frac{1}{4g^2}\l(\partial_\mu a_\nu-\partial_\nu a_\mu\r)^2\cr
\cl_{bf}&=&\lambda |\psi_\sigma|^2|\phi|^2\label{eq:lagrangian}
\end{eqnarray}
These terms are all allowed by the symmetries of the system, and thus will
generally arise in the coarse-graining process.
Note that the boson action is particle-hole symmetric, as appropriate for a
coarse-grained version of the slave rotor theory at half-filling.

At mean field level, we can ignore fluctuations of the gauge field $a_\mu$.
In the regime $m^2 > 0$, the charged rotor field
is gapped, so that $\langle \phi_i \rangle = 0$,
while the spinons form a gapless Fermi surface. This is the spin liquid
insulator. On the other hand,
the metallic phase at $m^2 < 0$ corresponds to a Bose-Einstein condensate
of rotors with $\langle \phi_i \rangle \neq 0$.  The spinon Fermi
surface now becomes an electron Fermi surface with
quasi-particle weight proportional to $|\langle \phi_i \rangle|^2$.
Therefore, the transition to the metal is tuned by the boson gap
parameter $m^2$, with $m^2 \propto U - U_c$. The
MIT, in this mean field theory, is thus
simply a Bose condensation of rotors with the spinon Fermi surface
being continuously connected to the electron Fermi surface, and hence it
is a continuous transition.

We now consider a full theory of this Mott transition obtained by analysing fluctuation
fluctuation corrections to all low-energy degrees of freedom
(spinons, rotors and gauge fields) in a self-consistent fashion. As the methods have already been elaborated in the 2D case in Ref. \onlinecite{senthil2d} we will be brief and mainly state the results.

\begin{figure}
\includegraphics[width=3.0 in]{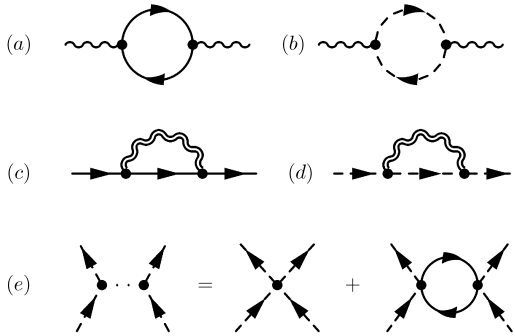}
 \vskip-0.1in
\caption{Dominant fluctuations.  The solid and dashed arrows represent fermions and bosons, respectively. {\bf (a)} and {\bf (b)} are the gauge
polarization functions arising from integrating out fermions and bosons, respectively. {\bf (c)} Spinon self-energy, where the double wavy line is the dressed gauge propagator.  {\bf (d)} Boson self-energy.  {\bf (e)} Residual bosonic interactions.
\label{fig:diag} } \vskip-0.1in
\end{figure}

We begin by considering the boson interaction term $u |\phi|^4$ together with the fermion-boson density interaction $\cl_{bf}$, in the absence of the gauge field $a_{\mu}$. (The gauge field is reinstated below). Upon integrating out fermions (see Fig.~\ref{fig:diag}(e)), this yields an effective boson-boson interaction
\be
\cl' =\! \int_{\begin{array}{c}\nu_n\nu'_n\omega_n\\ k,k',q \end{array}}\!\!\! {\widetilde u}(q,i\omega_n)\phi^\dagger_{k+q,\nu_n+\omega_n}\phi_{k,\nu_n} \phi^\dagger_{k'-q,\nu'_n-\omega_n}\phi_{k',\nu'_n}\nonumber
\ee
where ${\widetilde u}=u+N_0 \lambda^2  \pi_f (q,i\omega_n)$, $N_0$ is the fermionic density of states, and $\pi_f (q,i\omega_n)$
is the density-density polarization function of the fermions.
For $|q|\ll 2 k_f$, and in the limit $|\omega_n| \ll v_f q$, $\pi_f \to 2-\frac{|\omega_n|}{v_f q}$.
Under a naive rescaling of space
and time with dynamical critical exponent $z=1$, $\cl'$ is marginal by power 
counting at the critical fixed point of the bosonic theory, just like the 
conventional $u |\phi|^4$ term. The fate of this term can be 
determined by a standard perturbative renormalization group analysis. 
Equivalently
we generalize the theory to $N_b$ boson flavors, and consider the
$N_b=\infty$ limit of the theory.  In this limit, all diagrams can be summed 
exactly to yield a renormalized bosonic interaction
$u^{-1}_{\rm ren}(q,\omega)=
{\widetilde u}^{-1} (q,\omega)+\ln (\Lambda / \sqrt{\omega^2+q^2+m^2})$.
Near the QCP, the second term dominates, and $u_{\rm ren}\to 0$.  Hence, $\cl'$ is (dangerously) marginally irrelevant.

Now we reinstate the gauge field and first consider its renormalization  from integrating out the matter fields.  We work in the Coulomb gauge $\nabla\cdot {\textbf {\textit a}}=0$, in which
$a_0$ and ${\textbf {\textit a}}$ decouple in $\cl_g$.  Then, the longitudinal field $a_0$ is screened
by the gapless fermions, and therefore can be ignored.   On the other hand, the transverse part of ${\textbf {\textit a}}$
remains unscreened, but it is strongly renormalized by the matter fields, whose contribution dominates
over the bare gauge field action.
In the Random Phase Approximation (RPA), the renormalized inverse transverse gauge field propagator is
given by $D^{-1} ({\bf q},i\nu_n) = \Pi_f ({\bf q},i\nu_n) +\Pi_b ({\bf q},i\nu_n)$
where $\Pi_f$ and $\Pi_b$ are fermion and boson polarization functions.

The fermionic contribution, shown in Fig.~\ref{fig:diag}(a), is
$\Pi_f = \gamma_0\frac{|\nu_n|}{q}+\chi_0 q^2$.
The term proportional to $\gamma_0$ describes the Landau damping due to the gapless spinons
and $\chi_0$ is the diamagnetic susceptibility of the spinons.
This form can be further justified in the large $N_f$ limit, where
$N_f$ is the number of flavors of the fermions. The
corrections appearing in a $1/N_f$ expansion do not change this 
form\cite{kim94}.
The bosonic contribution shown in Fig.~\ref{fig:diag}(b) for $m^2 \ge 0$ (or $U \ge U_c$) is
$\Pi_b =\frac{q^2}{24\pi^2} \ln\frac{m_0}{(q^2+m^2)^{1/2}}$,
where $m_0$ is a non-universal constant, leading to
$\Pi_b = \frac{q^2}{24\pi^2} \ln\frac{m_0}{q}$ at the critical point $U=U_c$
and $\Pi_b \approx \frac{q^2}{24\pi^2} \ln\frac{m_0}{m}$ for $U > U_c$.
On the other hand, when $m^2 < 0$ or $U < U_c$,
$\Pi_b \approx \rho_s$ where $\rho_s \propto | \langle \phi \rangle |^2$ is the `superfluid stiffness' of
the bosons. Thus the bosons determine the different forms of the 
gauge field propagator in various regimes.

It is readily seen that the boson self-energy $\Sigma_b(\bq,\omega)$, given by Fig.~\ref{fig:diag}(d), 
acquires only analytic corrections in $\bq$ and $\omega$ and thus the bosons are not renormalized in an essential way by the gauge fluctuations. The arguments in Ref. \onlinecite{senthil2d} now imply that the boson sector is basically not affected by the gauge fluctuations in the scaling limit in the vicinity of
the critical point.

 The fermion self-energy $\Sigma_f$ shown in Fig.~\ref{fig:diag}(c) can be computed by
using the gauge field propagators obtained above. 
Near the spinon Fermi surface, $|{\bf k}| \approx k_f$, this leads to
$\Sigma_f \sim \omega \ln\ln 1 / |\omega|+i \frac{\pi}{2}\frac{|\omega|}{\ln 1/|\omega|}$
at the critical point $U=U_c$ and $\Sigma_f \sim \omega \ln {1 / |\omega|} + i {\pi \over 2} |\omega|$ in
the spin liquid phase for $U > U_c$.

Finally direct calculation shows that the $c_{\alpha} \rightarrow \phi + f_\alpha$ vertex is not singularly enhanced at low frequency. Thus we may calculate the singular part of the physical electron Greens function
by simply convolving the $\phi$ and $f_\alpha$ Greens functions.
In the metallic phase, we get the electron Green's function
${\cal G}_e = |\langle \phi \rangle|^2 {\cal G}_f = {Z_e \over \omega - \epsilon^*_p + i \delta}$
with $\epsilon^*_p \sim \frac{\epsilon_p}{\ln \ln 1/\rho_s}$ and $Z_e \sim {\rho_s \over \ln \ln {1/ \rho_s}}$.
Using $\rho_s \propto \left(U_c - U\right) \ln \frac{1}{U_c - U}$, the effective mass of the electrons $m^*_e/m_e \sim \ln \ln {1 / (U_c - U)}$
diverges very weakly. The quasi-particle weight $Z_e \sim \frac{\left(U_c - U \right)\ln \frac{1}{U_c - U}}{\ln\ln {1 / (U_c - U)}}$ diminishes as one
approaches the QCP.

Thus we see that there are only weak calculable logarithmic corrections due to fluctuations on top of the slave rotor mean field theory  
of the Mott transition in 3D. This enables us to make a number of firm predictions about the universal properties of the transition. 

{\it Scaling, Finite Temperature Phase Diagram, and Specific Heat}.---
As in 2D the finite $T$ crossovers will involve multiple energy scales originating in the different space-time scaling of the gauge field and the bosons.
Note that the mean-field transition of bosons at finite temperature, below which
$\rho_s \not = 0$, becomes a crossover in the full theory and represents the onset of
the charge coherence at this temperature scale.

The behavior of the charge excitations is governed by
$T \sim \omega \sim q$ scaling ($z=1$ theory). In either phase for $T$ above a crossover scale $T^* \sim 
|U- U_c|^{1/2}$, 
the charge bosons are in their quantum critical regime. In the insulator the scale $T^*$ may also be identified with the zero temperature charge gap.

The scaling behavior of the coupled spinon-gauge system is, however,  determined by $T \sim \omega \sim q^3$ scaling
($z=3$ theory) up to logarithmic corrections (in principle, one should use $\omega \sim q^3 \ln 1/q$
scaling). The spinon-gauge system thus emerges out of the quantum critical regime only at a scale $T^{**} \sim |U- U_c|^{\frac{3}{2}}$. 
This crossover scale is in principle visible in the specific heat 
which behaves as 
\begin{eqnarray}
\,\,\,\,\,\,\, C \approx \left\{
\begin{array}{l c l}
T \ln \ln 1/T  &  \ \ \ \ \  & T > |U-U_c|^{3/2} \\
\gamma_1 T \ln 1/T &  & T < (U -U_c)^{3/2} \\
\gamma_2 T & & T < (U_c-U)^{3/2}
\end{array}\right.\label{eq:specificHeat}
\end{eqnarray}
where $\gamma_1 \propto 1 / \ln {1 \over (U-U_c)}$ and
$\gamma_2 \sim \ln \ln {1 \over (U_c-U)}$. Note however that there only are rather weak changes in properties across this second 
crossover temperature $T^{**}.$

The crossover scale $T^*$ is, on the other hand, visible in the electrical conductivity which is readily measured in experiments. It can be obtained from the Ioffe-Larkin rule, which states that in a slave-particle formalism,
the electrical {\it resistivity} of the system is the sum of the resistivities of the fermions and the bosons, {\it i.e.} $\rho=\rho_f+\rho_b$.
For the fermions, we assume that the conductivity is dominated by elastic scattering due to impurities in the system, $\sigma_f=\sigma_{f,0}$.  The
bosonic conductivity, on the other hand, is  strongly temperature and pressure dependent.
Inside the Mott insulator, when the bosons are gapped ($T < T^*$),
the bosonic conductivity is activated $\sigma_b(T) \sim e^{-T^*/T}$.
On the other hand, inside the metal,
the bosons are condensed, and $\sigma_b(T\to 0) \to \infty$.
In the quantum critical regime,  standard scaling arguments give
$\sigma_b(T)\sim T\ln^2 \frac{1}{T}$ (the log corrections are due to the marginally irrelevant boson-boson interactions). As this goes to zero for small $T$, the boson contribution dominates
over that of the fermions due to the Ioffe-Larkin rule. 
Hence, the critical point itself is insulating.

These results for the conductivity are summarized in Fig.~\ref{fig:mf}(b) and (c).
Note that in the metallic phase near the critical point, $\sigma$ is not monotonic with temperature.
At high $T$, the system is in the (insulating) quantum critical fan, and conductivity decreases
with decreasing $T$.  It is only at low temperatures that the system realizes it is inside the metal,
and that conductivity begins to rise.  Hence, it is possible for a system to be metallic, and
yet display ``insulating" behavior at high temperatures.  The non-monotonicity becomes more
marked for cleaner samples, {\it i.e.} when $\sigma_{f,0}$ is large.
The experimental observation of such a conductivity would provide dramatic evidence for
the 3D Mott-metal transition discussed here.

{\it Mean-Field Theory for Hyper-Kagome Lattice.---}
We have shown that the bosons and spinons are only weakly affected by the gauge fluctuations near the quantum critical point.   This suggests that the mean-field theory can capture some essential features of the continuous Mott transition
in 3D. Encouraged by this observation, we specialize to the hyperkagome lattice
(of Na$_4$Ir$_3$O$_8$) and closely investigate the mean-field theory by approximating
the ground and excited states via direct products of spinon and rotor 
wave functions $|\Psi\rangle=|\Psi_\theta\rangle|\Psi_f\rangle$  \cite{Erhai}.
In the mean field analysis \cite{hyperkagomeMFT}, we indeed find a continuous 
quantum phase transition at $U_c \approx 6$ from
a spin-liquid insulator to a metallic phase.
The electron spectral function is given by,
\begin{eqnarray}
A_e (\bk,\omega) & = &
\sum_{\bq,\alpha} \frac{M_\alpha(\bk,\bq)}{2\Omega_\bq}\times 
\label{eq:spectral} \\
&\times& \left[(1+n(\Omega_\bq) -
f(\xi^\alpha_{\bk-\bq}))
\delta(\omega - \xi^\alpha_{\bk-\bq} - \Omega_\bq)  \right. \nonumber \\
&+& \left.
(n(\Omega_\bq) + f(\xi^\alpha_{\bk-\bq}))
\delta(\omega - \xi^\alpha_{\bk-\bq} + \Omega_\bq) \right]\nonumber
\end{eqnarray}
where $\xi^\alpha_\bk$ is the dispersion of the spinon band $\alpha=1\ldots 12$, $\Omega_\bq$ is the boson dispersion, and $n$ and $f$
are the Bose and Fermi distribution functions, respectively. Here, the 
`form factor' $M_\alpha(\bk-\bq)$ arises from the spinon wave function in 
band $\alpha$.
At the QCP, we assume the boson dispersion to be
given by $\Omega^2_\bq = 4 v^2_b (\sin^2(q_x/2)+\sin^2(q_y/2)
+\sin^2(q_z/2))$, so that $\Omega_\bq \approx v_b |\bq|$ for small momenta.
For illustrative purposes, we choose $v_b = 0.5$. Deep in the metallic
phase at $T=0$, where the bosons are fully condensed,
$A^{\rm (met)}_e(\bk,\omega)=
\sum_{\alpha} M_\alpha(\bk) \delta(\omega - \xi^\alpha_{\bk})$.

\begin{figure}
\includegraphics[width=3.4in]{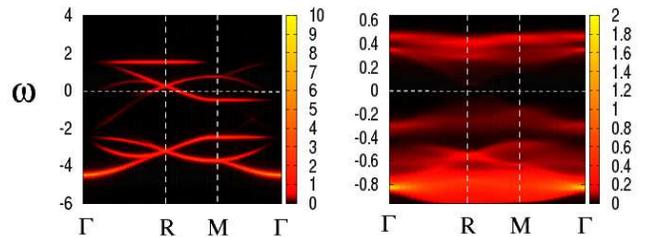}
\caption{Intensity plot of the electron spectral function along high symmetry directions
in the metallic phase (left) and at the critical point (right). Here
$\Gamma=(0,0,0)$, $R=(\pi,\pi,\pi)$, and $M=(\pi,\pi,0)$. $\omega=0$ marks the chemical potential.}
\vskip -0.2in
\label{fig:akw}
\end{figure}

Figure \ref{fig:akw} shows the $T=0$ spectral function along some
high symmetry directions of the Brillouin zone (BZ).
Deep in the metallic phase, the spectral function clearly shows twelve
bands (after accounting for the four-fold degeneracy of the
flat band at high energy).
Momentum variations of the `form factor' $M_\alpha(\bk)$ 
lead to changes in the intensity for the various bands as one traverses
the BZ. In particular, only the lowest 
band has nonzero intensity at the $\Gamma$-point.
As one approaches the QCP, the overall bandwidth 
gets suppressed due to the Hubbard $U$. In contrast to the metallic 
phase, the spectra at the QCP are considerably broadened 
due to rotor fluctuations. Finally, as evident from Fig. \ref{fig:akw},
there is significant suppression of spectral weight around the
chemical potential. This `pseudogap' is also reflected in the tunneling
density of states obtained by integrating $A^{\rm (crit)}(\bk,\omega)$
over all ${\bf k}$.  We find $N^{\rm (crit)}_{\rm tunn}(\omega)\sim \omega^2$ at low frequencies.

{\it Discussion.---}
In Ref.~\onlinecite{lawler08b}, it was shown that the specific heat of the insulating paramagnetic phase of Na$_4$Ir$_3$O$_8$ at not-too-low temperatures is well described by
a renormalized mean-field theory of a gapless spin liquid with a spinon Fermi surface.
Such success, however, may seem at odds with the fact that the gauge field fluctuations
would give rise to $C/T \sim \ln 1/T$ in the spin liquid regime while the renormalized
mean-field theory does not necessarily capture the gauge fluctuation effects  at finite temperatures. In light of the present work, it is possible
that the spin liquid phase
of Na$_4$Ir$_3$O$_8$ may already be near the QCP so that the temperature range
where the {\em bona fide} spin liquid phase can be seen is quite narrow. In this case,
the finite temperature behavior of the specific heat may be dominated by the C-SL
regime, leading to $C/T \sim \ln \ln 1/T$. This weak singularity would indeed be
consistent with the apparent success of the renormalized mean-field theory.
The recent discovery of the insulator-metal transition in Na$_4$Ir$_3$O$_8$
upon increasing pressure \cite{takagi08}
may also be consistent with the picture that the spin liquid
phase discovered earlier \cite{okamoto} is already close to the QCP. Furthermore,
our analysis suggests that
the gapless spin liquid ground state with a spinon Fermi surface itself is more likely to
occur due to the proximity to the Mott transition and the associated charge fluctuations
while it may face stiff competition from 
other possible phases deep inside the Mott insulator.
The confirmation of this picture would come from careful comparisons
between future experiments and the detailed predictions on the finite temperature
crossover phase diagram, specific heat, and conductivity presented
in this work. 

We thank Prof. H.~Takagi for helpful discussions and for sharing unpublished data with us.
This work was supported by CIFAR, CRC (DP, YBK), the NSERC of Canada (DP, AP, YBK), Sloan Foundation and Ontario ERA (AP), and
NSF Grant DMR-0705255 (TS).

\end{document}